\newcommand{\dst}{\Delta_{ST}(t)}
\newcommand{\drms}{\sigma_{ST}}
\newcommand{\dhf}{\Delta_{HF}(t)}
\newcommand{\dhfb}{\Delta_{HF}(t)}
\newcommand{\dhfrms}{\sigma_{HF}}
\newcommand{\dsob}{\Delta_{SO}}
\newcommand{\dso}{\Delta_{SO}}

\newcommand{\singsing}{| (1,1) S \rangle}
\newcommand{\sing}{| (0,2) S \rangle}
\newcommand{\Ss}{|S \rangle}
\newcommand{\trip}{| (1,1) T_+ \rangle}
\newcommand{\dbz}{\delta B_z}
\newcommand{\nd}{90^{\circ}}
\newcommand{\zd}{0^{\circ}}
\newcommand{\B}{\mathbf{B}}
\newcommand{\Osob}{\mathbf{\Omega}_{SO}}
\newcommand{\Oso}{\Omega_{SO}}

\newcommand{\dbperp}{\delta B_{\perp}}
\newcommand{\est}{\epsilon_{ST}}
\newcommand{\ta}[1]{\left\langle #1 \right\rangle}

\documentclass[%
reprint,amsmath,amssymb,aps,prl,superscriptaddress,
]{revtex4-1}

\usepackage{graphicx}
\usepackage{dcolumn}
\usepackage{bm}
\usepackage{epstopdf}

\begin{document}

\title{Quenching of dynamic nuclear polarization by spin-orbit coupling in GaAs quantum dots}

\author{John M. Nichol}

\author{Shannon P. Harvey}

\author{Michael D. Shulman}

\author{Arijeet Pal}

\affiliation{Department of Physics, Harvard University, Cambridge, MA, 02138, USA}

\author{Vladimir Umansky}

\affiliation{Braun Center for Submicron Research, Department of Condensed Matter Physics,
Weizmann Institute of Science, Rehovot 76100 Israel}

\author{Emmanuel I. Rashba}

\author{Bertrand I. Halperin}

\author{Amir Yacoby}

\affiliation{Department of Physics, Harvard University, Cambridge, MA, 02138, USA}

\begin{abstract}
The central-spin problem, in which an electron spin interacts with a nuclear spin bath, is a widely studied model of quantum decoherence~\cite{Chen2007}. Dynamic nuclear polarization (DNP) occurs in central spin systems when electronic angular momentum is transferred to nuclear spins~\cite{Abragam1978} and is exploited in spin-based quantum information processing for coherent electron and nuclear spin control~\cite{Foletti2009}. However, the mechanisms limiting DNP remain only partially understood~\cite{Chekhovich2013}. Here, we show that spin-orbit coupling quenches DNP in a GaAs double quantum dot~\cite{Petta2005}, even though spin-orbit coupling in GaAs is weak. Using Landau-Zener sweeps, we measure the dependence of the electron spin-flip probability on the strength and direction of in-plane magnetic field, allowing us to distinguish effects of the spin-orbit and hyperfine interactions.  To confirm our interpretation, we measure high-bandwidth correlations in the electron spin-flip probability and attain results consistent with a significant spin-orbit contribution. We observe that DNP is quenched when the spin-orbit component exceeds the hyperfine, in agreement with a theoretical model. Our results shed new light on the surprising competition between the spin-orbit and hyperfine interactions in central-spin systems.
\end{abstract}


\maketitle
Dynamic nuclear polarization occurs in many condensed matter systems, and is used for sensitivity enhancement in nuclear magnetic resonance~\cite{Gram2003} and for detecting and initializing solid-state nuclear spin qubits~\cite{Simmons2011}. DNP also occurs in two-dimensional electron systems~\cite{Wald1994} via the contact hyperfine interaction. In both self-assembled~\cite{Lai2006,Eble2006,Tartakovskii2007,Latta2009,Greilich2007} and gate-defined quantum dots~\cite{Ono2004,Foletti2009,Bluhm2010,Laird2007}, for example, DNP is exploited to create stabilized nuclear configurations for improved quantum information processing. Closed-loop feedback~\cite{Bluhm2010} based on DNP, in particular, is a key-component in one- and two-qubit operations in singlet-triplet qubits~\cite{Foletti2009,Shulman2012,Dial2013}.

Despite the importance of DNP, it remains unclear what factors limit DNP efficiency in semiconductor spin qubits~\cite{Chekhovich2013}. In particular, the relationship between the spin-orbit and hyperfine interactions~\cite{Stepanenko2012,Rudner2010,Neder2014} has been overlooked in previous experimental studies of DNP in quantum dots. In this work we show that spin-orbit coupling competes with the hyperfine interaction and ultimately quenches DNP in a GaAs double quantum dot~\cite{Petta2005,Shulman2012}, even though the spin orbit length is much larger than the interdot spacing. We use Landau-Zener (LZ) sweeps to characterize the static and dynamic properties of $\dst$, the splitting between the singlet $S$ and $m_s=1$ triplet $T_+$, and the observed suppression of DNP agrees quantitatively with a new theoretical model.

\begin{figure}
\includegraphics{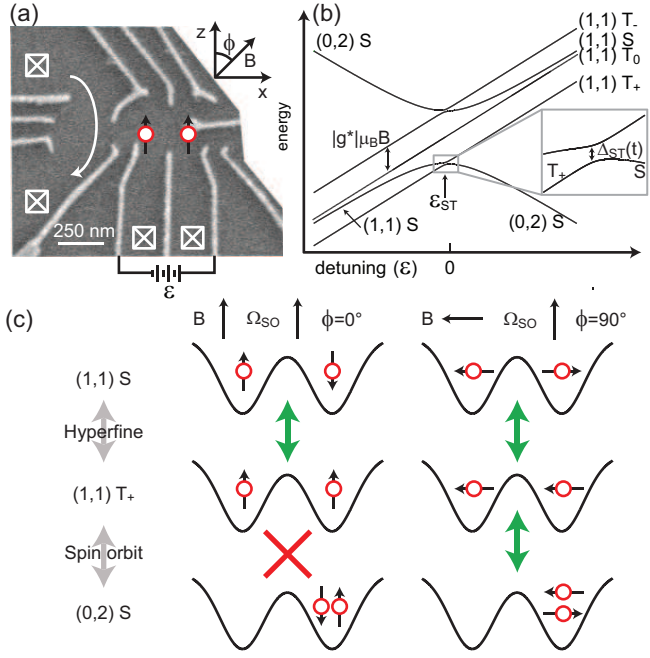}
\caption{\label{fig:1} Experimental setup. (a) Scanning electron micrograph of the double quantum dot. A voltage difference between the gates adjusts the detuning $\epsilon$ between the potential wells, and a nearby quantum dot on the left senses the charge state of the double dot. The gate on the right couples the double dot to an adjacent double dot, which is unused in this work. The angle between $\B$ and the $z$ axis is $\phi$. (b) Energy level diagram showing the two-electron spin states and zoom-in of the $S-T_+$ avoided crossing. (c) The hyperfine interaction couples $|(1,1)S\rangle$ and $|(1,1)T_+\rangle$ when the two dots are symmetric, regardless of the orientation of $\B$, and the spin-orbit interaction couples $|(0,2)S\rangle$ and $|(1,1)T_+\rangle$ when $\B$ has a component perpendicular to $\Osob =\Oso\hat{z}$, the effective spin-orbit field experienced by the electrons during tunneling.}
\end{figure}

\begin{figure}
\includegraphics{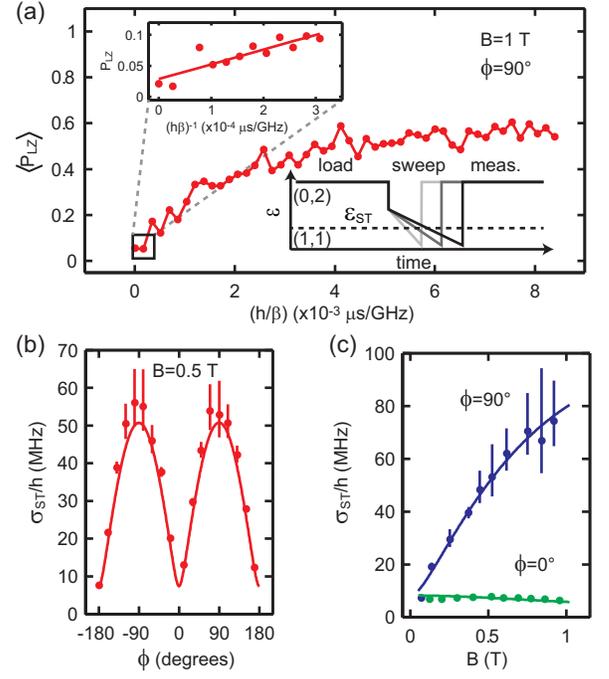}
\caption{\label{fig:2} Measurements of $\drms$. (a) Data for a series of LZ sweeps with varying rates, showing reduction in maximum probability due to charge noise. The horizontal axis is proportional to the sweep time. Upper inset: Data and linear fit for fast sweeps such that $0<\langle P_{LZ} \rangle<0.1$. Lower Inset: In a LZ sweep, a $\sing$ state is prepared, and $\epsilon$ is swept through $\est$ (dashed line) with varying rates. Here $h = 2 \pi \hbar$ is Planck's constant. (b) $\drms$ vs $\phi$ (dots) and simulation (solid line). (c) $\drms$ vs $B$ for $\phi=\zd$ and $\phi=\nd$ (dots) and fits to equation~(\ref{eq:delta})~(solid lines). Error bars are fit errors.}
\end{figure}

Figure~\ref{fig:1}(a) shows the double quantum dot used in this work~\cite{Petta2005,Shulman2012}. The detuning, $\epsilon$, between the dots determines the ground-state charge configuration, which is either (1,1) [one electron in each dot], or (0,2) [both electrons in the right dot] as shown in Fig.~\ref{fig:1}(b). To measure $\dst$, the electrons are initialized in $|(0,2)S\rangle$, $\epsilon$ is swept through the $S-T_+$ avoided crossing at $\epsilon=\est $, and the resulting spin state is measured [Fig.~\ref{fig:2}(a)]. In the absence of noise, slow sweeps cause transitions with near-unity probability. For large magnetic fields, however, we find maximum transition probabilities of approximately 0.5. This reduction is a result of rapid fluctuations in the sweep rate arising from charge noise (see Supplementary Information). Even in the presence of noise, however, the average LZ probability $\langle P_{LZ}(t) \rangle $ can be approximated for fast sweeps as $\frac{2 \pi \langle |\dst|^2 \rangle }{\hbar \beta}$ (see Supplementary Information). Here $\langle \cdots \rangle$ indicates an average over the hyperfine distribution and charge fluctuations, and $\beta=d(E_{S}-E_{T_+})/dt$ is the sweep rate, with $E_S$ and $E_{T_+}$ the energies of the $S$ and $T_+$ levels. To accurately measure  $\drms \equiv \sqrt{\langle |\dst|^2 \rangle}$, we therefore fit $\langle P_{LZ} \rangle$ vs $\beta^{-1}$ to a straight line for values of  $\beta$ such that $0<\langle P_{LZ} \rangle<0.1$.  [Fig.~\ref{fig:2} (a)]. 

We first measure $\drms$ vs $\phi$ at $B = 0.5$ T [Fig.~\ref{fig:2}(b)], where $\phi$ is the angle between the magnetic field $\B$ and the z axis [Fig.~\ref{fig:1}(a)].  $\drms$ oscillates between its extreme values at $\zd$ and $\nd$ with a periodicity of $180^{\circ}$. Fixing $\phi = \zd$ and varying $B$, we find that $\drms$ decreases weakly with with $B$, but when $\phi=\nd$, $\drms$ increases steeply with $B$, reaching values greater than 10 times that for $\phi=\zd$, as shown in Fig.~\ref{fig:2}(c). 

We interpret these results by assuming that both the hyperfine and spin-orbit interactions contribute to $\dst$ and by considering the charge configuration of the singlet state at $\est$ [Figs.~\ref{fig:1}(b) and (c)]. The matrix element between $S$ and $T_+$ can be written as $\dst=\dhfb+\dsob$. $\dhf=g^* \mu_B \dbperp (t)$ is the hyperfine contribution, which arises from the difference in perpendicular (relative to $B$) hyperfine field, $\dbperp(t)$, between the two dots~\cite{Taylor2007}. (In the following, we set $g^* \mu_B = 1$.) $\dhf$, which is a complex number, couples $\singsing$ to $\trip$ when the two dots are symmetric.  $\dso$ is the spin-orbit contribution, which arises from an effective magnetic field $\Osob=\Oso \hat{z}$ experienced by the electron during tunneling~\cite{Stepanenko2012}.  Only the component of $\Osob \perp \B$ causes an electron spin flip. $\dsob$ therefore couples $\sing$ to $\trip$ when $\phi \neq \zd$, and $\Oso$ is proportional to the double-dot tunnel coupling~\cite{Stepanenko2012}, which is 23.1 $\mu$eV here.  At $\est$, the singlet state $|S \rangle$ is a hybridized mixture: $| S \rangle= \cos\theta\singsing+\sin\theta\sing$, where the singlet mixing angle $\theta$ approaches $\pi/2$ as $B$ increases (see Supplementary Information). Taking both $\theta$ and $\phi$ into account, we write~\cite{Stepanenko2012}
\begin{eqnarray}
\dst&=&\dhfb+\dsob \nonumber \\
&=& \dbperp(t)\cos\theta+\Oso\sin\phi\sin\theta\label{eq:delta}.
\end{eqnarray}

The data in Fig.~\ref{fig:2}(b) therefore reflect the dependence of $\dst$ on $\phi$ in equation~(\ref{eq:delta}).  The data in Fig.~\ref{fig:2}(c) reflect the dependence of $\dst$ on $\theta$. As $B$ increases, $\theta$ also increases, and $\Ss$ becomes more $\sing$-like, causing $\dhf$ to decrease. When $\phi=\zd$, $\dso=0$ for all $B$, but when $\phi=\nd$, $\dso = \Oso\sin\theta$, and  $\drms$ increases with $B$. Fitting the data in Fig.~\ref{fig:2}(c) allows a direct measurement of the spin-orbit and hyperfine couplings (see Supplementary Information). We find $\sqrt{\langle |\dbperp^2 (t)| \rangle} = 34 \pm 1$ neV and $\Oso = 461 \pm 10$ neV, corresponding to a spin-orbit length $\lambda_{SO}\approx 13$ $\mu$m~\cite{Stepanenko2012}, in good agreement with previous estimates in GaAs~\cite{Petta2010,Nowack2007,Shafiei2013}. 

We further verify that $\dst$ contains a significant spin-orbit contribution by measuring the dynamical properties of $P_{LZ}(t)$. A key difference between the spin-orbit and hyperfine components is that $\dsob$ is static, while $\dhfb$ varies in time because it arises from the transverse Overhauser field, which can be considered a precessing nuclear polarization in the semiclassical limit~\cite{Taylor2007}. To distinguish the components of $\dst$ through their time-dependence, we develop a high-bandwidth technique to measure the power spectrum of $P_{LZ}(t)$.

Instead of measuring the two-electron spin state after a single sweep, $\epsilon$ is swept twice through $\est$ with a pause of length $\tau$ between sweeps [Fig.~\ref{fig:3}(a)] (See Supplementary Information). Assuming that St\"uckelberg oscillations rapidly dephase during $\tau$~\cite{Petta2010,Dial2013}, and after subtracting a background and neglecting electron spin relaxation, the time-averaged triplet return probability is proportional to $R_{PP}(\tau)\equiv\langle P_{LZ}(t)P_{LZ}(t+\tau)\rangle$, the autocorrelation of the LZ probability [Fig.~\ref{fig:3}(b)]. Taking a Fourier-transform therefore gives $S_{P}(\omega)$, the power spectrum of $P_{LZ}(t)$ [Figs.~\ref{fig:3}(c) and \ref{fig:3}(d)]. For $P_{LZ}(t)\ll 1$, $P_{LZ}(t) \propto |\dst|^2$, so $S_{P}(\omega) \propto S_{|\Delta_{ST}|^2}(\omega)$, the power spectrum of $|\dst|^2$. This two-sweep technique allows us to measure the high-frequency components of $S_P(\omega)$, because the maximum bandwdith is not limited by the quantum dot readout time. 

\begin{figure}
\includegraphics{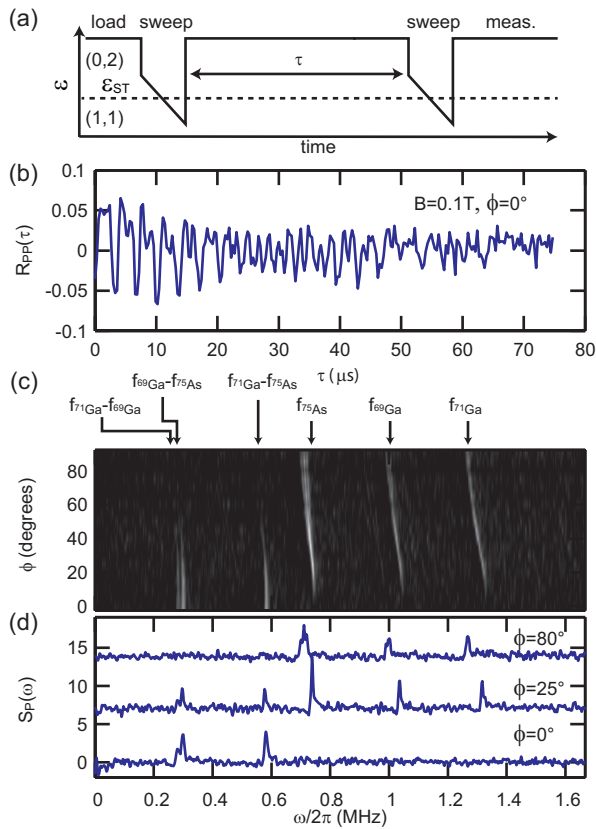}
\caption{\label{fig:3} Correlations and power spectrum of $P_{LZ}(t)$. (a) Pulse sequence to measure $R_{PP}(\tau)$ using two LZ sweeps. (b) $R_{PP}(\tau)$ for $\phi=0^{\circ}$ and $B=0.1$ T. The data extend to $\tau=200$ $\mu$s, but for clarity are only shown to 75 $\mu$s here. (c) $S_P(\omega)$ vs $\phi$ obtained by Fourier-transforming $R_{PP}(\tau)$. At $\phi=\zd$, the differences between the nuclear Larmor frequencies are evident, but for $|\phi|>\zd$, the absolute Larmor frequencies appear, consistent with a spin-orbit contribution to $\drms$. The reduction in frequency with $\phi$ is likely due to the placement of the device slightly off-center in our magnet (see Supplementary Information). (d) Line cuts of $S_P(\omega)$ at $\phi = \zd$, 25$^{\circ}$, and 80$^{\circ}$. }
\end{figure}

Because it arises from the precessing transverse nuclear polarization, $\dhf$ contains Fourier components at the Larmor frequencies of the $^{69}$Ga, $^{71}$Ga, and $^{75}$As nuclei in the heterostructure, i.e., $\dhfb = \sum_{\alpha=1}^{3}\Delta_{\alpha}e^{2 \pi i f_{\alpha}t+\theta_{\alpha}}$, where $\alpha = 1,2$, or 3 indexes the three nuclear species, and the $\theta_{\alpha}$ are the phases of the nuclear fields.  Without spin-orbit interaction, $|\dst|^2=|\sum_{\alpha=1}^{3}\Delta_{\alpha}e^{2 \pi i f_{\alpha}t+\theta_{\alpha}}|^2 $ contains only Fourier components at the differences of the nuclear Larmor frequencies. With a spin-orbit contribution, however, $|\dst|^2=|\dsob+\dhfb|^2$ contains cross-terms like $\dso\Delta_{\alpha}e^{2 \pi i f_{\alpha}t+\theta_{\alpha}}$ that give $|\dst|^2$ Fourier components at the absolute Larmor frequencies. A signature of the spin-orbit interaction would therefore be the presence of the absolute Larmor frequencies in $S_P (\omega)$ for $\phi \neq \zd$~\cite{Dickel2014}.  

Figure~\ref{fig:3}(b) shows $R_{PP}(\tau)$ measured with $B=0.1$ T and $\phi=\zd$. Figure~\ref{fig:3}(c) shows $S_P(\omega)$ for $\zd \leq \phi \leq \nd$. At $\phi=\zd$, only the differences between the Larmor frequencies are evident, but as $\phi$ increases, the absolute nuclear Larmor frequencies appear, as expected for a static spin-orbit contribution to $\dst$. These results, including the peak heights, which reflect isotopic abundances and relative hyperfine couplings, agree well with simulations (see Supplementary Information). 

\begin{figure}
\includegraphics{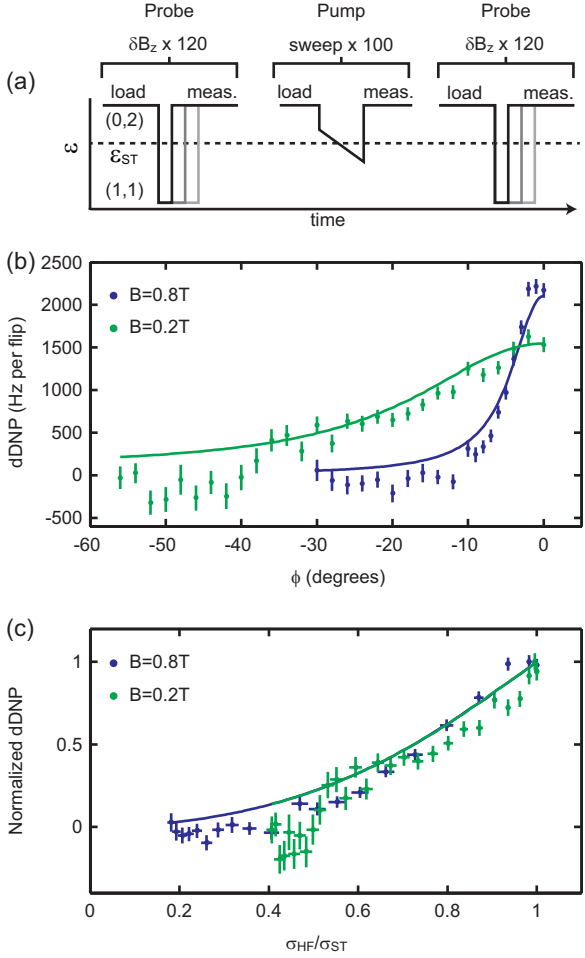}
\caption{\label{fig:4} DNP quenching by spin-orbit coupling. (a) Protocol to measure DNP. $\dbz$ is measured before and after 100 LZ sweeps by evolving the electrons around $\dbz$.  (b) dDNP vs $\phi$ at fixed $\ta{P_{LZ}}=0.4$ for $B=0.8$ T and $B=0.2$ T and theoretical curves (solid lines). dDNP is suppressed for $|\phi| > 0$ because of spin-orbit coupling. (c) Data and theoretical curves for fixed $\ta{P_{LZ}}$ collapse when normalized and plotted vs $\dhfrms/\drms$. Vertical error bars are statistical uncertainties, and horizontal error bars are fit errors.}
\end{figure}

Having established the importance of spin-orbit coupling at the $S-T_+$ crossing, we next investigate how the spin-orbit interaction affects DNP.  Previous research has shown that repeated LZ sweeps through $\est$ increase both the average and  differential nuclear longitudinal polarization in double quantum dots~\cite{Foletti2009}. However, the reasons for left/right symmetry breaking, which is needed for differential DNP (dDNP), and the factors limiting DNP efficiency in general are only partially understood. Here, we measure dDNP precisely by measuring $\dbz$, the differential Overhauser field, using rapid Hamiltonian learning strategies~\cite{Shulman2014} before and after 100 LZ sweeps to pump the nuclei with rates chosen such that $\ta{P_{LZ}}=0.4$ (see Supplementary Information)  [Fig.~\ref{fig:4}(a)]. 
 
Figure~\ref{fig:4}(b) plots the change in $\dbz$ per electron spin flip for $B = 0.2$ T and $B=0.8$ T for varying $\phi$. In each case, the dDNP decreases with $|\phi|$. Because the spin-orbit interaction allows electron spin flips without corresponding nuclear spin flops, dDNP is suppressed as $|\dso| = |\Oso\sin\phi\sin\theta|$ increases  with $|\phi|$. The reduction in dDNP occurs more rapidly at 0.8 T because $\dso$ is larger at 0.8 T than at 0.2 T. We gain further insight into this behavior by plotting the data against $\dhfrms/\drms$, where $\dhfrms \equiv \sqrt{\langle |\dhf|^2\rangle}$ [Fig~\ref{fig:4}(c)]. Plotted in this way, the two data sets show nearly identical behavior, suggesting that the size of the hyperfine interaction relative to the total splitting primarily determines the DNP efficiency.  

Based on theoretical results and experimental data, to be presented elsewhere, we expect that the dDNP should be proportional to the total DNP, with a constant of proportionality that depends on $B$, but not $\beta$ or $\phi$. We therefore explain our measurements of dDNP using a theoretcal model in which we have computed the average angular momentum $\langle \delta m \rangle$ transfered to the ensemble of nuclear spins following a LZ sweep as:
\begin{eqnarray}
\langle \delta m \rangle \propto \dhfrms^2  \left\langle \frac{P'_{LZ}(\Delta_{ST})}{|\Delta_{ST}|} \right\rangle \label{eq:deltam},
\end{eqnarray}
where $P'_{LZ}(\Delta_{ST})$ is the derivative of the LZ probability with respect to the magnitude of the splitting. (See Supplementary Information for more details.) Neglecting charge noise, we have the usual Landau-Zener formula~\cite{Petta2010}
\begin{eqnarray}
P_{LZ}(\Delta_{ST})=1-\exp\left( - \frac{2 \pi |\Delta_{ST}|^2}{\hbar \beta}\right), \label{eq:LZ}
\end{eqnarray}
and equation~(\ref{eq:deltam}) reduces to 
\begin{eqnarray}
\langle \delta m \rangle \propto \dhfrms^2 \frac{2 \pi}{\hbar \beta}  \left\langle 1- P_{LZ}\right\rangle \label{eq:deltam_2}.
\end{eqnarray}
The data in Figs.~\ref{fig:4}(b) and (c) can therefore be understood in light of equation~(\ref{eq:deltam_2}) because as the splitting $\drms$ increases with $|\phi|$, the sweep rate $\beta$ was also increased to maintain a constant $\langle P_{LZ}\rangle$. Because the hyperfine contribution $\dhfrms$ is independent of $\phi$, $\langle \delta m \rangle$ therefore decreases.  The data collapse in Fig.~\ref{fig:4}(c) can also be understood from equation~(\ref{eq:deltam_2}), assuming a constant splitting and fixed probability. In this case, $\beta \propto |\Delta_{ST}|^2$, as follows from equation~(\ref{eq:LZ}), and hence $\langle \delta m \rangle \propto \dhfrms^2/ |\Delta_{ST}|^2$. Measurements with fixed rate $\beta$ also exhibit a similar suppression of dDNP (see Supplementary Information). In this case $\langle P_{LZ} \rangle$ increases with $|\phi|$, because of the increasing spin-orbit contribution to $\drms$, and according to equation~(\ref{eq:deltam_2}), $\langle \delta m \rangle$ therefore decreases.  

The two theoretical curves in Figs.~\ref{fig:4}(b) and (c) are calculated using equation~(\ref{eq:deltam_2}) multiplied by fitting constants $C$, which are different for the two fields, and agree well with the data. As discussed in the Supplementary Information, we do not expect charge noise to modify the agreement between theory and data in Figs.~\ref{fig:4}(b) and (c) beyond the experimental accuracy. Interestingly, the peak dDNP is less at $B=0.2$ T than at $B=0.8$ T, perhaps because the electron-nuclear coupling becomes increasingly asymmetric with respect to the center of the quantum dots at higher fields~\cite{Brataas2011}. Finally, the peak dDNP value also approximately agrees with a simple calculation (see Supplementary Information) based on measured properties of the double dot.

In summary, we have used LZ sweeps to measure the $S-T_+$ splitting in a GaAs double quantum dot. We find that the spin-orbit coupling dominates the hyperfine interaction and quenches DNP for a wide range of magnetic field strengths. A misalignment of $\B$ to $\Osob$ by only $5^{\circ}$ at $B=1$ T can reduce the DNP rate by a factor of two, and DNP is completely suppressed for a misalignment of $15^{\circ}$. The techniques developed here are directly applicable to other quantum systems such as InAs or InSb nanowires and SiGe quantum wells, where the spin-orbit and hyperfine interactions compete. On a fundamental level, our findings suggest avenues of exploration for improved $S-T_+$ qubit operation~\cite{Petta2010} and underscore the importance of the spin-orbit interaction in the study of nuclear dark states~\cite{Gullans2010,Brataas2012} and other mechanisms that limit DNP efficiency in central-spin systems.

\begin{acknowledgments}
 This research was funded by the United States Department of Defense, the Office of the Director of National Intelligence, Intelligence Advanced Research Projects Activity, and the Army Research Office grant W911NF-11-1-0068. S.P.H was supported by the Department of Defense through the National Defense Science Engineering Graduate Fellowship Program. This work was performed in part at the Harvard University Center for Nanoscale Systems (CNS), a member of the National Nanotechnology Infrastructure Network (NNIN), which is supported by the National Science Foundation under NSF award No. ECS0335765.
\end{acknowledgments}

\bibliography{spin_orbit}

\end{document}


\title{Supplementary Information for\\
Quenching of dynamic nuclear polarization by spin-orbit coupling in GaAs quantum dots}

\author{John M. Nichol}

\author{Shannon P. Harvey}

\author{Michael D. Shulman}

\author{Arijeet Pal}

\affiliation{Department of Physics, Harvard University, Cambridge, MA, 02138, USA}

\author{Vladimir Umansky}

\affiliation{Braun Center for Submicron Research, Department of Condensed Matter Physics,
Weizmann Institute of Science, Rehovot 76100 Israel}

\author{Emmanuel I. Rashba}

\author{Bertrand I. Halperin}

\author{Amir Yacoby}

\affiliation{Department of Physics, Harvard University, Cambridge, MA, 02138, USA}


\pacs{}

\maketitle

\section{Measuring $\drms$}
Here we describe the fitting procedure to extract $\drms$. The experimentally measured quantity is the average triplet occupation probability $\langle P_T \rangle$, which we interpret as the average Landau-Zener (LZ) probability $\langle P_{LZ} \rangle$, at the end of a sweep. Here $\langle \cdots \rangle$ indicates an average over the hyperfine distribution and charge fluctuations for the same nominal sweep parameters. We calibrate the rate $\beta=d(E_S-E_{T_+})/dt$ using the spin-funnel technique~\cite{Petta2010} and assume a linear change in the $S-T_+$ splitting near the avoided crossing. 

\begin{figure}
\includegraphics{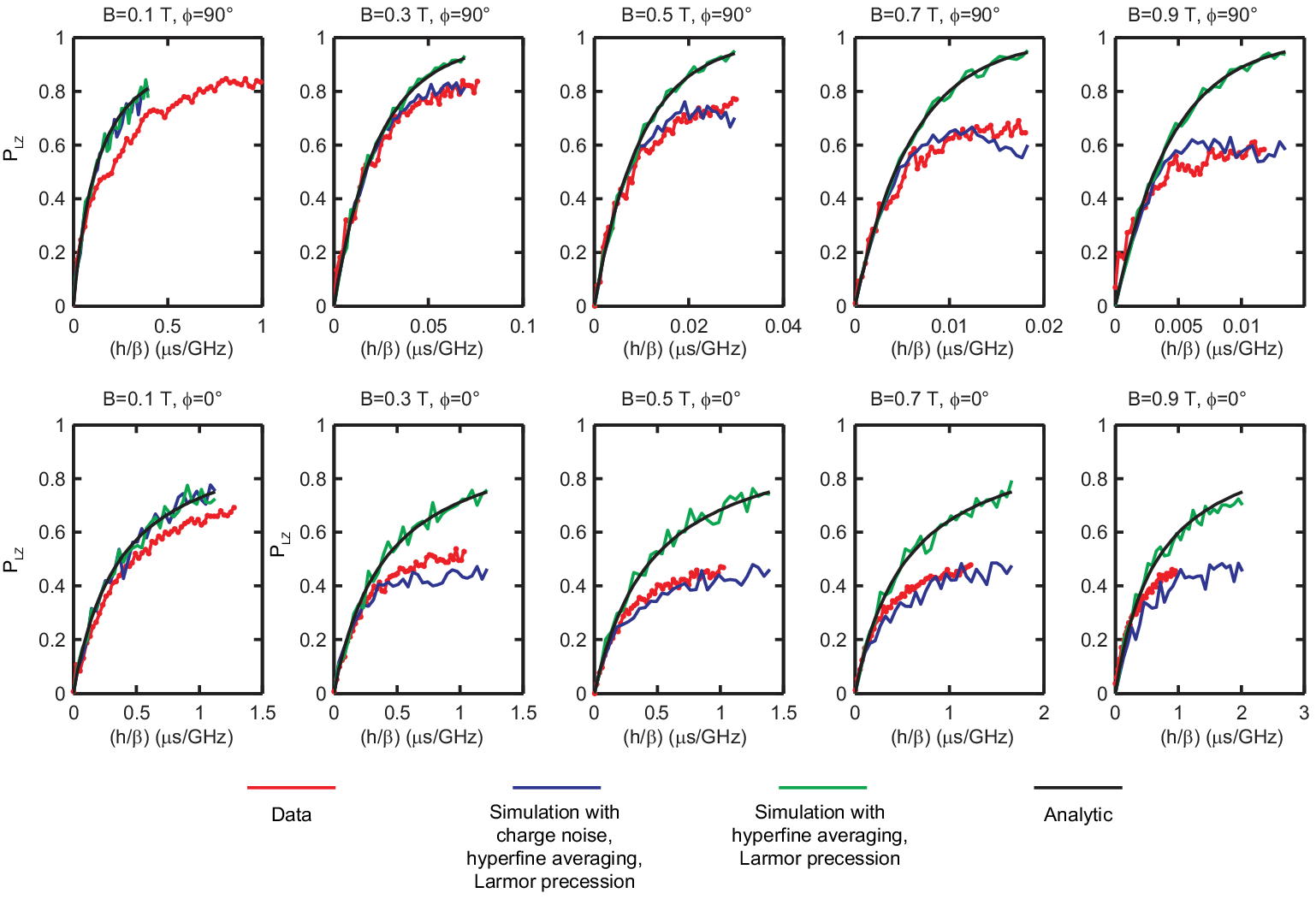}
\caption{\label{supfig:noise} Comparison of LZ data and simulations. Each panel shows data and simulations for a different magnetic field strength and orientation. Red curves are experimental data for a series of LZ sweeps with varying rates.  Blue curves are simulated data including charge noise, hyperfine averaging, and nuclear Larmor precession for the calculated value of the splitting corresponding to the red curves. Green curves are simulated data with hyperfine averaging and Larmor precession for the same value of the splitting as the blue curves. Black curves are calculated via equation~(\ref{supeq:exact}) using the same value of the splitting. In all panels, the y axis is $\langle P_{LZ} \rangle$, and the x axis is $h/\beta~(\mu$s/GHz). Here $h=2 \pi \hbar$ is Planck's constant.}
\end{figure}

$\dhf$ varies in time because of the nuclear Larmor precession and statistical fluctuations in the magnitude of the nuclear polarizations. We argue that both types of hyperfine fluctuations occur on time scales much longer than LZ transitions and can be treated as quasi-static. In typical experiments, the $S-T_+$ splitting is swept through approximately 5 GHz in less than 1 $\mu$s. For splittings of order 10 MHz, the total time spent near the avoided crossing is less than 10 ns, which is much faster than the nuclear Larmor period at 1 T, roughly 100 ns. Furthermore, during 1 $\mu$s, the nuclear polarization diffuses by approximately 7 kHz~\cite{Shulman2014}, which is 3 orders of magnitude smaller than $\dhfrms$. We therefore assume that the splitting is constant during a single sweep. Numerical simulations discussed below also support the hypothesis that nuclear Larmor precession does not significantly affect $\ta{P_T}$ for the sweep rates used here [Fig.~\ref{supfig:noise}]. 

In the absence of hyperfine or charge fluctuations, the probability for a transition is given by the LZ formula: $P_{LZ}(t)=1- \exp(-2\pi |\dst|^2/ (\hbar \beta))$~\cite{Shevchenko2010}. Neglecting high-frequency charge noise, the exact form of the LZ probability averaged over the hyperfine distribution can be computed. Let the total splitting be $\Delta_{ST}=\Delta_{HF}+\dso$. We take $\dso$ to be the constant, real spin-orbit part and $\Delta_{HF}$ the complex hyperfine contribution. Assuming that the real and imaginary parts of  $\Delta_{HF}$ ($\dhx$ and $\dhy$, respectively) are Gaussian-distributed around zero such that the root-mean-square hyperfine splitting is $\dhfrms$, the probability distribution for the splitting to have magnitude $\de=|\de_{ST}|$ is

\begin{eqnarray}
p(\de)&=&\frac{1}{\pi \dhfrms^2} \int_{-\infty}^{\infty} d \dhx \int_{-\infty}^{\infty} d \dhy ~ e^{-\frac{{\dhx}^2+{\dhy}^2}{\dhfrms^2}}\delta\left(\de-\sqrt{(\dso+\dhx)^2+{\dhy}^2}\right)\\
&=&\frac{2 \de}{\dhfrms^2}e^{-\frac{\de^2+\dso^2}{\dhfrms^2}}I_0(2\de\dso/\dhfrms^2)\label{supeq:dist},
\end{eqnarray}
where $I_0$ is the zeroth-order modified Bessel function of the first kind. Note that when $\dso=0$, equation~(\ref{supeq:dist}) reduces to the familiar distribution $p(\de)=\frac{2 \de}{\dhfrms^2}e^{-\de^2/\dhfrms^2}$~\cite{Neder2014}.   Integrating the LZ probability over this distribution yields the average LZ probability $\langle P_{LZ} \rangle$:
\begin{eqnarray}
\langle P_{LZ} \rangle&=& \int_0^{\infty} d\de \left( 1-\exp\left(-\frac{2 \pi \de^2}{\hbar \beta}\right) \right) p(\de) \\
&=& 1-Q\exp\left(-\frac{2 \pi \dso^2}{\hbar \beta}Q\right)\label{supeq:exact},
\end{eqnarray}
with 
\begin{eqnarray}
Q=\frac{1}{1+\frac{2 \pi \dhfrms^2}{\hbar \beta}}.
\end{eqnarray}
Note that this result agrees with another derivation~\cite{Dickel2014}. Note also that to leading order in $\beta^{-1}$, $\langle P_{LZ} \rangle \approx 2 \pi \left(\dso^2 + \dhfrms^2 \right)/ \hbar \beta$.

The average triplet return probability $\ta{P_T}$ may be modified due to effects of charge noise on the defining gates or in the two-dimensional electron gas itself. High-frequency charge noise in double quantum dots has recently been identified as a major source of decoherence~\cite{Dial2013}. In the current setting, corrections to $\ta{P_T}$ should occur, because charge fluctuations lead to time-dependent variations in $S-T_+$ detuning $E_S-E_{T_+}$, on top of the linear time-dependence due to the prescribed sweep rate $\beta$. Additionally, charge fluctuations can add noise to the off-diagonal coupling $\dst=\dbperp(t)\cos\theta+\Oso\sin\phi\sin\theta$, because the singlet mixing angle $\theta= \tan^{-1}\left(\frac{\epsilon + \sqrt{\epsilon^2+4t^2}}{2t}\right)$ depends on $\epsilon$. (Here $t=23.1$ $\mu$eV is the double-dot tunnel coupling.) As discussed below, however, the noise in $\Delta_{ST}$ should have much less effect than the detuning noise for the magnetic fields studied here.  

We observe that for high magnetic fields and slow sweeps, the maximum LZ probability falls to 0.5 as shown in Fig.~\ref{supfig:noise}. It was previously noted that strong detuning noise can have such an effect~\cite{Kayanuma1984}. To confirm that charge noise causes the probability reduction, we have performed Monte Carlo simulations of the Schr\"odinger equation for symmetric double dots undergoing LZ sweeps, including the effects of wide-band charge noise, nuclear Larmor precession, and averaging over the hyperfine distribution. The results of the simulations and experimental data are shown in Fig.~\ref{supfig:noise}. We generate random charge noise with power spectrum $14 \times 10^{-14}$ $\frac{\textnormal{V}^2}{\textnormal{Hz}} \left(\frac{1 \textnormal{Hz}}{f}\right)^{0.7}$ for  $f<\textnormal{1 GHz}$, and 0 otherwise. We generate the Fourier transform of the charge noise time record by picking the amplitude corresponding to the chosen power spectrum and a random phase for each frequency $f$ in the desired range. We then perform an inverse Fourier transform to obtain the charge noise time record. The spectrum we chose corresponds to a noise amplitude of 3 nV/$\sqrt{\textnormal{Hz}}$ at $f=1$ MHz, which is approximately the measured level of charge noise in the double dot used here. Note that we have  extrapolated the $f^{-0.7}$ frequency dependence that was previously measured to $f=1$ MHz in ref.~\cite{Dial2013} up to $f=1$ GHz in these simulations. However, one expects the results to be most sensitive to noise in the range of 10-100 MHz, corresponding to the size of the splitting. The $\epsilon$-dependent Hamiltonian used in these simulations was 

\begin{align}
H(\epsilon)=
\begin{pmatrix}
\frac{\epsilon}{2}-B & \dbperp(t)\cos\theta+\Oso\sin\phi\sin\theta \\
\dbperp^*(t)\cos\theta+\Oso\sin\phi\sin\theta & -\frac{1}{2}\sqrt{\epsilon^2+4t^2}
\end{pmatrix}
\end{align}
in the $\{|T_+\rangle, | S\rangle\}$ basis. Linear $\epsilon$ sweeps through the $S-T_+$ crossing $\est = \frac{B^2-t^2}{B}$ were used in the simulation to replicate the actual experiments. For each strength and orientation of the magnetic field, $\theta$ was calculated at $\est$ using the measured tunnel coupling, and the fitted values of the spin-orbit and hyperfine couplings from the main text were used to compute the splitting. We assumed a lever arm of 10 to convert the voltage noise on the quantum dot gates to $\epsilon$ noise.
 
The simulated LZ curves with charge noise agree well with the data as shown in Fig.~\ref{supfig:noise}.  The same simulations including averaging over the hyperfine distribution and nuclear Larmor precession, but without charge noise, show very little reduction in probability compared with the analytic result, equation~(\ref{supeq:exact}), supporting the hypothesis that charge noise is responsible for most of the observed probability reduction. A key feature in these experiments is the decreasing maximum probability with increasing magnetic field. We can understand that this trend occurs because the effect of charge
noise on the Landau Zener probability is controlled by the fluctuation in
the energy splitting $\delta E(\epsilon)$ produced by a given fluctuation
in the detuning $\eps$, which is proportional to $\frac{d E(\eps)}{d \eps}|_{\eps = \est}$. Since $E (\eps) = \frac{\epsilon}{2}-B+\frac{1}{2}\sqrt{\epsilon^2+4t^2}$, the magnitude of $\frac{d E(\eps)}{d \eps}|_{\eps = \est}$ increases sharply with increasing magnetic field.

\begin{figure}
\includegraphics{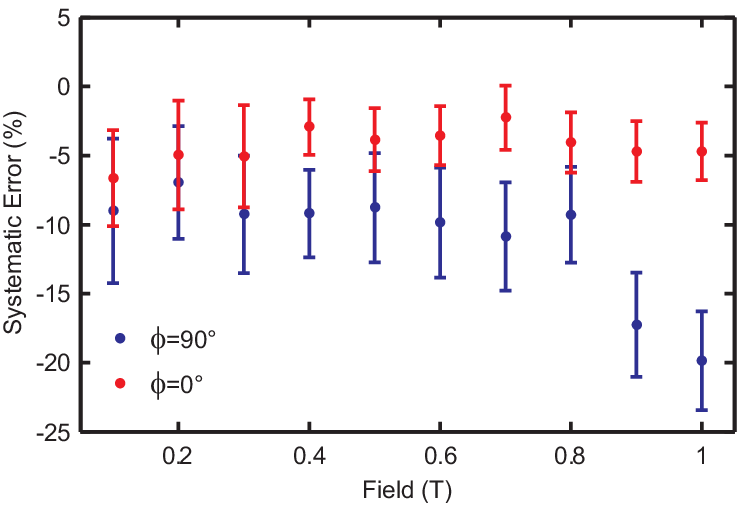}
\caption{\label{supfig:fiterr} Fitting error. We compute the fitting error by simulating $\langle P_{LZ} \rangle$ for the calculated splitting at each of the magnetic field configurations in the presence of charge noise.  The simulated $\langle P_{LZ} \rangle$ vs $\beta^{-1}$ is fitted to a straight line for $0<\langle P_{LZ} \rangle<0.1$, and the fitted value of the splitting is subtracted from the value chosen for the simulation. The difference is then divided by the simulated value of the splitting. Error bars are fit errors.}
\end{figure}

Even in the presence of noise, however, the average LZ probability in the limit of fast sweeps is still $2\pi |\dst|^2/ \hbar \beta$, which is identical to the leading order behavior of the usual LZ formula, as shown in section 3.1 of ref.~\cite{Kayanuma1984}. Replacing the LZ formula in equation~(\ref{supeq:exact}) by its leading order behavior, and performing the integration over the quasi-static distribution gives $\langle P_{LZ} \rangle \approx 2 \pi \left(\dso^2 + \dhfrms^2 \right)/ \hbar \beta$. Such a result can be understood because the effect of detuning noise is reduced on short time scales. Figure~\ref{supfig:noise} demonstrates this idea because the analytic curves deviate significantly from the data for $\langle P_{LZ} \rangle \gtrsim 0.2$, but for $0 < \langle P_{LZ} \rangle < 0.1$, the analytic results agrees well with the data. Based on additional simulations, we estimate the systematic error in the deduced value of $\drms$ as obtained by fitting measured values of $\ta{P_{LZ}}$ to a straight line for values of $\beta$ such that $0<\ta{P_{LZ}}<0.1$ to be small for most of the experimental conditions as shown in Fig.~\ref{supfig:fiterr}.

We note that the coupling $\dst = \dbperp(t)\cos\theta+\Oso\sin\phi\sin\theta$ depends on $\epsilon$ through the singlet mixing angle $\theta$. This dependence means that during a LZ sweep, the coupling $\dst$ varies both due to the linear $\eps$ sweep as well as charge noise. We estimate that $\frac{d E(\eps)}{d\eps} \geq 40  \frac{d \drms}{d \eps}$ for the fields studied here. We therefore expect detuning fluctuations to be the dominant noise source. Furthermore, when $|E(\eps)|<\drms$, $\drms$ changes by only a few percent during the sweep and is likely not a significant source of error in the measurement of $\dst$.  Additionally, we note that the simulations in Fig.~\ref{supfig:noise}, which include $\eps$-depending coupling, demonstrate that the fitting procedure described above allows an accurate measurement of $\drms$. Finally, we have also performed additional simulations, taking into account the measured values of $E(\eps)$, which deviate slightly from the values predicted by assuming a constant tunnel coupling, and we observe no significant change in our results.

\section{Direction of $\Osob$}
The double quantum dot axis is aligned within $\approx 5^{\circ}$ of either the $[\bar{1}10]$ or $[110]$ axes of the crystal, but we do not know which. In the later case, both the Rashba and Dresselhaus spin-orbit fields are aligned with the z axis, and their magnitudes add~\cite{Stepanenko2012}. In the former case, the Rashba and Dresselhaus contributions are also aligned with the z axis, but their magnitudes subtract. The techniques used here could be employed to distinguish the Rashba and Dresselhaus spin-orbit contributions by measuring $\drms$ with double quantum dots fabricated on different directions with respect to the crystal axes.

\section{Fitting $\drms$ vs B and $\phi$}
We fit the data in Fig. 2(c) in the main text to a function of the form $\drms=\sqrt{\dso^2 \sin^2 \theta \sin^2 \phi+\dhfrms^2 \cos^2 \theta }$, with $\dso$ and $\dhfrms$ as fit parameters. The singlet mixing angle $\theta$ is computed by assuming that the $(1,1)$ and $(0,2)$ singlet branches are a two-level system with constant tunnel coupling, as discussed above. 

$\dso$ is held at 0 when fitting data for $\phi=\zd$ to determine the hyperfine coupling. We also exclude data points for $B < 0.2$ T in the fit, as the hyperfine contribution appears to decrease at very low fields. We determine the spin orbit length using equation (28) of Ref. \cite{Stepanenko2012}, where the spin-orbit field is computed as $\Oso=\frac{4t}{3}\frac{\lambda_{DQD}}{\lambda_{SO}}$, where $\lambda_{DQD} \approx 200$ nm is the interdot spacing, and $\lambda_{SO}$ is the spin-orbit length. The simulation in Fig. 2(b) in the main text is generated using the same equation with the fitted values of the $\dso$ and $\dhfrms$.

\section{Measuring $R_{PP}(\tau)$}
Here we derive the triplet return probability after two consecutive LZ sweeps with a pause of length $\tau$ in between. In experiments, both sweeps were in the same direction, and $\eps$ was held in the $(0,2)$ region between sweeps, as shown in Fig. 3(a) in the main text. Suppose the first LZ sweep takes place at time $t$ with probability $P_{LZ}(t)$. The probability for the two electrons to be in the $T_+$ state is $P_{LZ}(t)$, while the probability to be in the $S$ state is $1-P_{LZ}(t)$. Then, the detuning is quickly swept into the $(0,2)$ region. Here, electron spin dephasing occurs rapidly, and there is very little $T_+$ occupation in thermal equilibrium because the $S$ and $T_+$ states are widely separated in energy. Thus, after a wait of length $\tau$, but before the second sweep, the triplet population is $P_{LZ}(t)e^{-\tau/T_1}$, and the singlet population is $1-P_{LZ}(t)e^{-\tau/T_1}$, where $T_1$ is the electron relaxation time. After the second sweep, the triplet occupation probability is
\begin{eqnarray}
P_T(t+\tau) &=& \left(1-P_{LZ}(t)e^{-\tau/T_1}\right)P_{LZ}(t+\tau)+P_{LZ}(t)e^{-\tau/T_1}\left( 1-P_{LZ}(t+\tau)\right)\\
&=& -2P_{LZ}(t)P_{LZ}(t+\tau)e^{-\tau/T_1}+P_{LZ}(t+\tau)+P_{LZ}(t)e^{-\tau/T_1}. \label{eq:Rpp}
\end{eqnarray}
The second and third terms in equation~(\ref{eq:Rpp}) vary slowly with $\tau$. These terms are found by fitting the measured triplet probability to an exponential with an offset and are subtracted. When $T_1 \gg \tau$, relaxation can be neglected, and the predicted time-averaged signal is $\langle P_T(t+\tau) \rangle \propto R_{PP}(\tau)$, where $R_{PP}(\tau) \equiv \langle P_{LZ}(t)P_{LZ}(t+\tau) \rangle$, the autocorrelation of the LZ probability. When $\phi = \zd$, $T_1 \gg \tau_{max}$ = 200 $\mu$s, where $\tau_{max}$ is the largest value of $\tau$ measured. The shortest relaxation time $T_1 \approx 100$ $\mu$s in these experiments time occurs when $\phi = \nd$, which is  consistent with spin-orbit-induced relaxation \cite{Scarlino2014}. 

The effect of $T_1$ relaxation is to multiply the measured correlation by an exponentially-decaying window, which reduces the spectral resolution of the Fourier transform, but does not shift the frequency of the observed peaks. We expect statistical fluctuations in the amplitude of the hyperfine field to affect the spectrum in a similar way, although we expect this effect to be less than that of electron relaxation. The raw data, [Fig. 3(b) in the main text]  consisting of 667 points (each a result of two sweeps with a 40 $\%$ chance of a LZ transition) spaced by 300 ns, were zero-padded to a size of 1691 points to smooth the spectrum, and a Gaussian window with time constant 150 $\mu$s was applied to reduce the effects of noise and ringing from zero-padding before Fourier transforming. 

\begin{figure}
\includegraphics{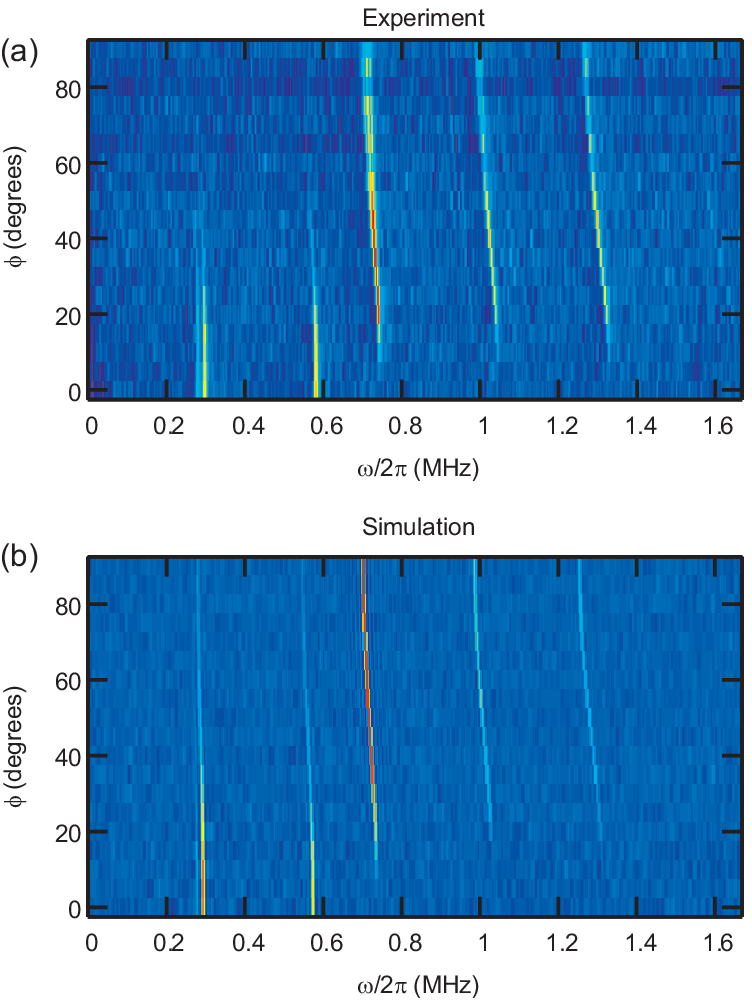}
\caption{\label{supfig:spectrum}Simulations of $S_P(\omega)$. (a) Experimental data. (b) Theoretical simulation taking into account known sweep rates, nuclear magnetic resonance frequencies, hyperfine couplings, and a 4.4$\%$ reduction in field in the x direction. The expected frequencies at $B=0.1$ T are $f_{^{69}Ga}=1.0248$~MHz, $f_{^{71}Ga}=1.302$~MHz, and $f_{^{75}As}=0.7315$~MHz.}
\end{figure}

The magnetic resonance frequencies in Fig. 3(c) appear to decrease with $\phi$. The inhomogeneity of the x-coil in our vector magnet is 1.6 $\%$ at 0.6 cm offset from the center. Thus, the field could easily be reduced by more than 3 $\%$ for a misplacement of the sample by 1 cm from the magnet center. We have simulated the data in Fig. 3(c) in the main text based on the measured hyperfine and spin-orbit couplings and the known sweep rates. Assuming a 4.4~$\%$ reduction in the field from the x-coil, we obtain good agreement between theory and experiment [Figs. \ref{supfig:spectrum}(a) and (b)]. 

We argued in the main text that only the difference frequencies should appear in the spectrum  $S_P (\omega)$ without spin-orbit coupling by considering the time-dependence of $|\dst|^2$ and because $S_P (\omega) \propto S_{|\Delta_{ST}|^2}(\omega)$ when $P_{LZ}(t) \ll 1$. Since $P_{LZ}(t)$ contains only even powers of $|\dst|$, $S_P (\omega)$ can generally be expressed in terms of differences of the resonance frequencies, but will not contain the absolute frequencies in the absence of spin-orbit coupling, regardless of the value of $P_{LZ}(t)$.

\section{Derivation of nuclear polarization change $\ta{\delta m}$}
Here we derive equations 2 and 4 in the main text. Let $\Delta_{ST} = \Delta_{SO}+\Delta_{HF}$ where $\Delta_{SO}$ is real and 
\begin{eqnarray}
\Delta_{HF}=\sum_j \lambda_j I_j^+,
\end{eqnarray}
where $I_j^+$ is the raising operator for the $j^\textnormal{th}$ nuclear spin, and the $\lambda_j$ are individual coupling constants. We assume that there are many nuclear spins, so that each coupling constant is small. Also,
\begin{eqnarray}
\dhfrms^2\equiv \langle |\Delta_{HF}^2|\rangle = \frac{2}{3}I(I+1)\sum_j \lambda_j^2 = \frac{5}{2}\sum_j \lambda_j^2,
\end{eqnarray}
where $I=\frac{3}{2}$ is the spin of the nuclei, and the angular brackets refer to an average over the distribution of nuclear spins. 

We pick one of the nuclear spins, $j$, and we wish to compute $\langle \delta m_j \rangle$, the mean value of the change in $I_j^z$ after one sweep. Let $P_{LZ}(\Delta_{ST})$ be the probability of an $S-T_+$ transition for a fixed value of $\Delta_{HF}$. Clearly, $P_{LZ}$ depends on $|\Delta_{ST}|$. We calculate $\delta m_j$ as follows. Write 
\begin{eqnarray}
\Delta_{ST}=a + b e^{i \theta_j},
\end{eqnarray}
where $a$ includes the contributions of spin orbit and of all nuclei other than the nucleus $j$, and the second term represents the contribution (of order $\lambda_j$) from nucleus $j$.  According to equation (31) of Ref.~\cite{Neder2014}, the value of $\delta m_j$ for this configuration should be given by
\begin{eqnarray}
\delta m_j = \frac{1}{2 \pi} \oint d\theta_j P_{LZ}(\Delta_{ST}) \frac{d \varphi}{ d \theta_j},
\end{eqnarray}
where $\varphi=\arctan(\textnormal{Im}(\Delta_{ST})/\textnormal{Re}(\Delta_{ST}))$ specifies the orientation of $\Delta_{ST}$ in the complex plane. Without loss of generality, we may suppose that $a$ is real. Then we have, ignoring terms that are higher order in $b/a$,
\begin{eqnarray}
\frac{d \varphi}{ d \theta_j} &=& \frac{b}{a} \cos\theta_j\\
P_{LZ}(\Delta_{ST})&=& P_{LZ}(a)+b P'_{LZ}(a)\cos\theta_j\\
\delta m_j&=&\frac{b^2}{2a}P'_{LZ}(a),
\end{eqnarray}
where $P'_{LZ}(a)$ is the derivative of $P_{LZ}(a)$ with respect to $a$. Averaging over nuclear configurations, we obtain
\begin{eqnarray}
\langle \delta m_j \rangle = \langle b^2 \rangle \left\langle \frac{P'_{LZ}(a)}{2a} \right\rangle,
\end{eqnarray}
with $\langle b^2 \rangle = (5/2)\lambda_j^2$. In the case of no charge noise, we have 
\begin{eqnarray}
P_{LZ}(\Delta_{ST})=1-\exp\left( - \frac{2 \pi |\Delta_{ST}|^2}{\hbar \beta}\right) \label{supeq:LZ},
\end{eqnarray} 
so
\begin{eqnarray}
\frac{P'_{LZ}(a)}{2a} = \frac{2 \pi}{\hbar \beta}(1-P_{LZ}(a)) \label{supeq:pprim}
\end{eqnarray}
and
\begin{eqnarray}
\langle \delta m_j \rangle = \frac{2 \pi}{\hbar \beta} \langle b^2 \rangle \langle 1 - P_{LZ}(a) \rangle. 
\end{eqnarray}
Finally, we sum over all nuclear spins and make the replacement $a \approx |\Delta_{ST}|$, obtaining
\begin{eqnarray}
\langle \delta m\rangle=  \frac{2 \pi}{\hbar \beta}  \dhfrms^2  \langle 1 - P_{LZ}(\Delta_{ST}) \rangle. \label{supeq:dm}
\end{eqnarray} 
The collapse demonstrated in Fig. 4(c) in the main text can be understood from equation~(\ref{supeq:dm}), assuming constant $\Delta_{ST}$ and fixed probability. In this case, $\beta \propto |\Delta_{ST}|^2$ from equation~(\ref{supeq:LZ}), and hence $\langle \delta m \rangle \propto \dhfrms^2/ |\Delta_{ST}|^2$.

In the case of a fixed splitting, equation~(\ref{supeq:dm}) reduces to 
\begin{eqnarray}
\langle \delta m\rangle=  \frac{2 \pi}{\hbar \beta}  \dhfrms^2  \exp \left( {-\frac{2 \pi |\Delta_{ST}|^2}{\hbar \beta}} \right).\label{supeq:fixed} 
\end{eqnarray}
In equation~(\ref{supeq:fixed}), $\langle \delta m\rangle \to 0$ for both $\beta \to 0$ and $\beta \to \infty$. In practice however, experiments necessarily average over the hyperfine distribution. Thus, using equation~(\ref{supeq:exact}) with $\dso = 0$ to compute $\langle 1- P_{LZ}(\Delta_{ST}) \rangle$, we have 
\begin{eqnarray}
\langle \delta m\rangle &=&  \frac{2 \pi}{\hbar \beta}  \dhfrms^2  Q\\
&=&    \frac{\frac{2 \pi \dhfrms^2}{\hbar \beta}  }{1+\frac{2 \pi \dhfrms^2}{\hbar \beta}}. \label{supeq:dmhf}
\end{eqnarray}
According to equation~(\ref{supeq:dmhf}), in the limit of slow sleeps, where $\beta \to 0$, $\langle \delta m\rangle \to 1$, and in the limit of fast sweeps, where $\beta \to \infty$, $\langle \delta m\rangle \to 0$, as expected.

The theory curves in Figs. 4(c) and (d) in the main text were generated by computing equation~(\ref{supeq:dm}). For each field angle $\phi$, the parameters $\theta$, $\dso$, and $\dhfrms$ were calculated using the fitted values of the spin-orbit and hyperfine couplings as well as the measured tunnel coupling. Equation~(\ref{supeq:exact}) was then solved using the calculated parameters to find the rate $\beta$ such that $\langle P_{LZ}\rangle$ = 0.4. In order to compare with data on the dDNP rate, the theoretical curves for $\ta{\delta m}$ were multiplied by fitting constants $C$, which are different for the two curves. As explained in the main text, and further discussed below, we expect the ratio between the dDNP rate and $\ta{\delta m}$ to depend on the magnetic field but to be independent of the sweep rate. 

\begin{figure}
\includegraphics{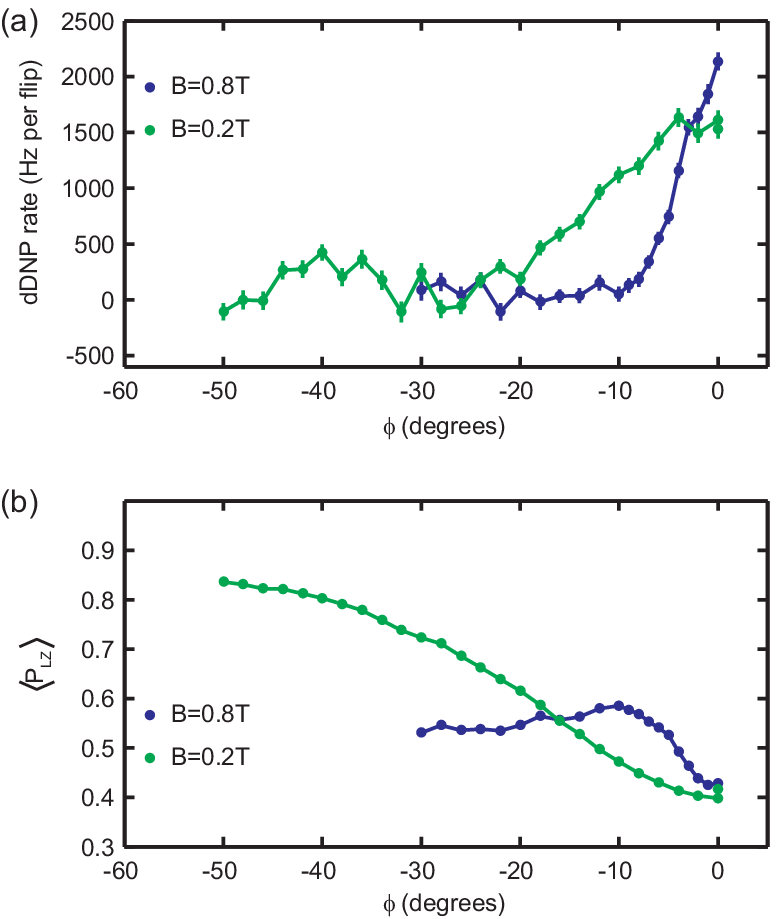}
\caption{\label{supfig:fixed_rate} DNP quenching with fixed sweep rate. (a) dDNP vs $\phi$ at $B=0.2$ T and $B=0.8$ T. For each field, the sweep rate $\beta$ was chosen to give $\ta{P_{LZ}}=0.4$ at $\phi=\zd$ and then was held constant for $\phi \neq \zd$. (b) As $|\phi|$ increases, $\drms$ increases. As a result, $\ta{P_{LZ}}$ also increases and DNP is suppressed, according to equation~(\ref{supeq:dm}). Error bars are statistical uncertainties. Lines between points serve as a guide to the eye.}
\end{figure}

Data taken at fixed sweep rate $\beta$ also show a suppression of DNP, as shown in Fig.~\ref{supfig:fixed_rate}(a). In this case, $\ta{P_{LZ}}$ increases with $|\phi|$ because of spin-orbit coupling [Fig.~\ref{supfig:fixed_rate}(b)], and $\ta{P_{LZ}}$ therefore increases, causing $\ta{\delta m}$ to decrease, according to equation~(\ref{supeq:dm}).

\begin{figure}
\includegraphics{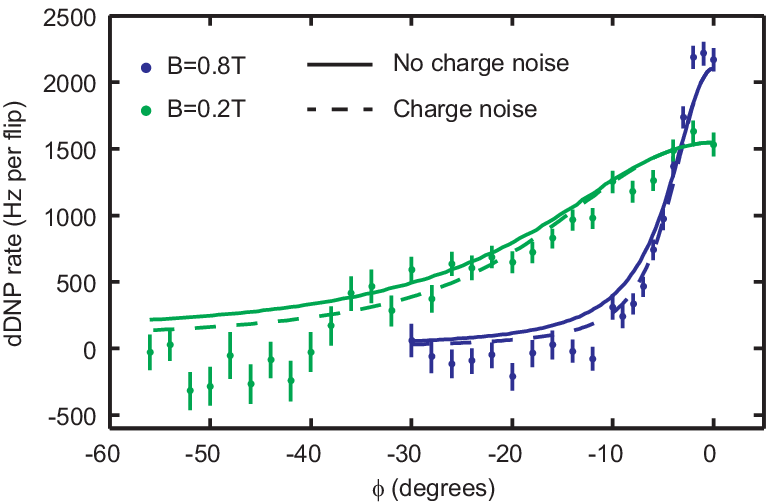}
\caption{\label{supfig:chrgnoise} The effect of charge noise on dDNP. The data and solid lines are the same as in Fig. 4 in the main text, and the dashed lines are the theoretical estimates for dDNP in the presence of charge noise. The dashed and solid lines are normalized to the same values at $\phi = \zd$. Error bars are statistical uncertainties.}
\end{figure}

To address the effect of charge noise on dDNP, we recompute equation~(\ref{supeq:dm}) in the limit of strong noise using the results of Ref.~\cite{Kayanuma1984}, making the replacement $P(a)=\frac{1}{2}\left( 1-\exp\left(- \frac{4 \pi a^2}{\hbar \beta}\right) \right)$ for $P_{LZ}(a)$ both in the derivation leading to equation~(\ref{supeq:dm}) and in equation~(\ref{supeq:exact}) for the computation of $\beta$. The expected dDNP in the presence of strong noise is shown in Fig.~\ref{supfig:chrgnoise}, and it does not significantly deviate from the case without noise, at least at the level of the experimental accuracy.

\section{Measuring $\delta B_z$}
We measure $\delta B_z$ by first initializing the double dot in the $\sing$ state and then separating the electrons by rapidly changing $\eps$ to a large negative value~\cite{Petta2005}. When the electrons are separated, the exchange energy is negligible, and the magnetic field gradient $\dbz$ drives oscillations between $| S \rangle$ and $| T_0 \rangle$. In our experiments, we measure the two-electron spin state for 120 linearly increasing values of the separation time. The resulting single-shot measurement record is thresholded, zero padded, and Fourier transformed. The frequency corresponding to the peak in the resulting Fourier transform is chosen as the value of $\delta B_z$. This technique is related to a previously described rapid Hamiltonian estimation technique~\cite{Shulman2014}.

\section{Expected DNP rate}
In this section we give a simple calculation to explain the value of the peak ($\phi = \zd$) dDNP rate, as shown in Fig. 4 of the main text. Additional measurements were carried out to measure the pumping rate of the sum hyperfine field, $(B_r + B_l)/2$, where $B_r$ and $B_l$ denote the longitudinal hyperfine fields in the right and left dots. This rate was determined by measuring the location of $\est$ before and after a series of LZ sweeps to polarize the nuclei at $B=0.2$ T. We observe that the sum field is pumped roughly twice as efficiently as the difference field, $\dbz=B_r-B_l$. Setting $(\dot{B_r}+\dot{B_l})/2=2(\dot{B_r}-\dot{B_l})$, where $\dot{B}_{l(r)}$ indicates the pumping rate of the left(right) dot, we have $\dot{B_l} = (3/5) \dot{B_r}$, meaning that the left dot is pumped $3/5$ as often as the right dot. Under these conditions, the average gradient builds up at a rate (per electron spin flip) of $(\dot{B_r}-\dot{B_l})/(\dot{B_r}+\dot{B_l})$ that is only 1/4 the rate that would occur if nuclear spin flips occurred in only one dot. 

To determine the expected change in $\dbz$, we require the approximate number of spins overlapped by the electronic wave function in the double dot. We have measured the inhomogeneous dephasing time of electronic oscillations around $\dbz$ and find $T_2^* = 18$ ns~\cite{Petta2005}. This dephasing time corresponds to a rms value of the gradient $\sigma_{\dbz}\equiv \sqrt{\langle |\dbz|^2 \rangle}=h/\left(|g^*| \mu_B \sqrt{2} \pi T_2^* \right)=2$ mT, where $h$ is Planck's constant. The total number of spins overlapped by the wavefunction is $N=\left( h_1 /\sigma_{\dbz}\right)^2\approx3\times10^6$, where $h_1 = 4.0$~T~\cite{Taylor2007}. If all nuclear spins were fully polarized, then the dots would experience a hyperfine field of $h_0=5.3$~T~\cite{Taylor2007}, and if the nuclear spins in the two dots were fully polarized in opposite directions, the gradient would be $2 h_0$. Therefore, the expected change in the gradient per electron spin flip, corresponding to a change in nuclear angular momentum of $\hbar$, is $\frac{2 \pi}{\hbar} \times \frac{2 |g^*| \mu_B h_0}{2I (N/2)} =  12$~kHz, where $I=3/2$ is the nuclear spin. The average dDNP under actual conditions is 1/4 of this value, or 3~kHz, in reasonable agreement with our observations. In addition, we note the reasonable agreement between the measured value of $\sigma_{\dbz}=2$~mT and the root-mean-square hyperfine gap $ \sqrt{\langle |\dbperp (t)|^2 \rangle} \approx 34$ neV/ $( |g^*| \mu_B) =  1.5$ mT.

\bibliography{spin_orbit_SI}